\documentclass{article}

\usepackage{arxiv}

\usepackage[utf8]{inputenc} 
\usepackage[T1]{fontenc}    
\usepackage{hyperref}       
\usepackage{url}            
\usepackage{booktabs}       
\usepackage{amsfonts}       
\usepackage{nicefrac}       
\usepackage{microtype}      
\usepackage{lipsum}		
\usepackage{graphicx}
\usepackage{natbib}
\usepackage{doi}
\usepackage{threeparttablex}
\usepackage{sistyle}
\SIthousandsep{,}

\usepackage{import}

\def\bz{{\mathbf z}}
\def\bZ{{\mathbf Z}}

\def\btheta{{\boldsymbol{\theta}}}
\def\bthetahat{\widehat{\boldsymbol{\theta}}}
\def\blambda{{\boldsymbol{\lambda}}}
\def\blambdahat{\widehat{\boldsymbol{\lambda}}}
\usepackage{amsthm}

\usepackage{booktabs}
\usepackage{amsmath}
\newcommand{\pkg}[1]{{\fontseries{m}\fontseries{b}\selectfont #1}}

\title{Extrapolation before imputation reduces bias when imputing censored covariates}

\author{ \href{https://orcid.org/0000-0001-5380-2427}{\includegraphics[scale=0.06]{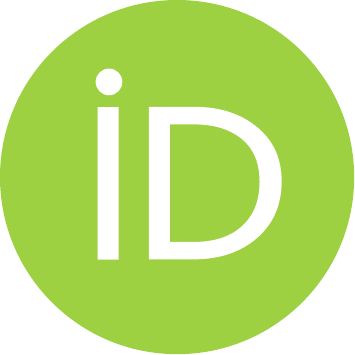}\hspace{1mm}Sarah C.~Lotspeich} \\
	Department of Statistical Sciences\\
	Wake Forest University\\
	Winston-Salem, NC 27109 \\
	\texttt{lotspes@wfu.edu} \\
	\And
	Tanya P.~Garcia \\
	Department of Biostatistics\\
	University of North Carolina at Chapel Hill\\
	Chapel Hill, NC 27599\\
	\texttt{tpgarcia@email.unc.edu} \\
}

\renewcommand{\shorttitle}{Extrapolation before imputation reduces bias when imputing censored covariates}

\hypersetup{
pdftitle={Escaping the trap},
pdfsubject={q-bio.NC, q-bio.QM},
pdfauthor={Sarah C.~Lotspeich, Tanya P.~Garcia},
pdfkeywords={Adaptive quadrature, Breslow's estimator, Conditional mean imputation, Huntington's disease, Time to diagnosis, Trapezoidal rule},
}

\begin{document}
\maketitle

\begin{abstract}
Modeling symptom progression to identify informative subjects for a new Huntington’s disease clinical trial is problematic since time to diagnosis, a key covariate, can be heavily censored. Imputation is an appealing strategy where censored covariates are replaced with their conditional means, but existing methods saw over 200\% bias under heavy censoring. Calculating these conditional means well requires estimating and then integrating over the survival function of the censored covariate from the censored value to infinity. To estimate the survival function flexibly, existing methods use the semiparametric Cox model with Breslow’s estimator, leaving the integrand for the conditional means (the estimated survival function) undefined beyond the observed data. The integral is then estimated up to the largest observed covariate value, and this approximation can cut off the tail of the survival function and lead to severe bias, particularly under heavy censoring. We propose a hybrid approach that splices together the semiparametric survival estimator with a parametric extension, making it possible to approximate the integral up to infinity. In simulation studies, our proposed approach of extrapolation then imputation substantially reduces the bias seen with existing imputation methods, even when the parametric extension was misspecified. We further demonstrate how imputing with corrected conditional means helps to prioritize patients for future clinical trials.
\end{abstract}

\keywords{Adaptive quadrature \and Breslow's estimator \and Conditional mean imputation \and Huntington's disease \and Time to diagnosis \and Trapezoidal rule}

\section{Introduction}
\label{sec:intro}

\subsection{Modeling the Progression of Huntington's Disease}

Prospective studies are common for genetically inherited diseases because, with genetic testing, researchers can identify at-risk subjects and follow their symptom development over time. Such studies are especially powerful for Huntington's disease, a genetically inherited neurodegenerative disease caused by unstable cytosine-adenine-guanine (CAG) repeats in the HTT gene \citep{HDCRG1993}. Huntington's disease is fully penetrant, so anyone with $\geq 36$ CAG is guaranteed to develop the disease. One such prospective study is the Neurobiological Predictors of Huntington's Disease (PREDICT-HD) \citep{paulsen2008detection}.

Modeling the progression of Huntington's disease using data from prospective studies like PREDICT-HD is appealing, for example, as we investigate experimental treatments designed to slow or delay symptoms. Models of how impairment (i.e., in daily, motor, and cognitive function) progresses relative to the time of clinical diagnosis can help identify subjects to recruit into clinical trials. Huntington's disease symptoms are most detectable in the few years immediately before and after a diagnosis, so subjects in this window of time would be ideal to test a new therapy in a clinical trial. 

However, Huntington's disease progresses slowly, with functional, motor, and cognitive decline spanning decades, so prospective studies often end before all at-risk subjects have met the diagnosis criteria. (A diagnosis is made when motor abnormalities are unequivocal signs of Huntington's disease \citep{Kieburtz1996}.) Therefore, the slow-moving nature of the disease leaves the key variable ``time to diagnosis'' right-censored among subjects who have yet to be diagnosed (i.e., their motor abnormalities will merit a diagnosis sometime after their last study visit, but exactly when is unknown). Thus, we face a pressing statistical challenge when investigating Huntington's disease progression: how to model the association between a fully observed outcome (impairment) and a randomly right-censored covariate (time to diagnosis).

\subsection{Imputing a Censored Covariate}

Inspired by missing data techniques, one appealing strategy is conditional mean imputation, where we replace all right-censored times to diagnosis with their conditional means \citep{AtemMatsoukaZimmern2019,AtemEtAl2017,AtemSampeneGreene2019}. This conditional mean imputation ensures that the imputed time to diagnosis is realistic (i.e., after the last study visit) and adjusts for other variables that may influence time to diagnosis (e.g., CAG repeat length). (Conditional mean imputation could be adopted in a single or multiple imputation framework. For simplicity, we focus on single imputation; however, multiple imputation would encounter the same challenges and could be corrected in the same ways that we introduce.) The conditional mean for a right-censored value is the expected time to diagnosis given that it must happen after the censored value (the last study visit) and additional covariates. In theory, this expected time to diagnosis can be anywhere from the last study visit to infinity, so computing it involves an integral over this range.

As we will discuss in Section~\ref{sec:imp}, calculating the conditional means involves integrating over the conditional survival function (the integrand) of the censored covariate up to infinity. Typically, this function relies on a step function (in this case, Breslow's estimator), which is well-defined up to the largest uncensored covariate value but not beyond that. If there are covariate values beyond the largest uncensored one, Breslow's estimator will carry forward the last estimated survival, but this is unrealistic in practice. In Huntington's disease studies, at-risk people who are not yet undiagnosed will be diagnosed eventually. Importantly, this step function leaves the integrand not well defined beyond the observed covariate values and to infinity, so more accurate quadrature alone will not improve the estimation of the conditional means. (This is ``typical'' because nonparametric or semiparametric estimators are often chosen because of their distribution-free robustness but they rely on step functions; a parametric estimator would already be defined up to infinity.)

Existing approaches to conditional mean imputation use the trapezoidal rule to compute the integral over Breslow's estimator from the censored value to the largest uncensored value in the data \citep{AtemMatsoukaZimmern2019,AtemEtAl2017,AtemSampeneGreene2019}. Specifically, they define partitions based on the observed covariate values and their corresponding survival estimates,  
relying on the data to define the integral's upper bound and ending the final partition at the largest observed covariate value. Thus, for the trapezoidal rule approximation over Breslow's estimator to hold in this improper integral case, the largest observed covariate value in the data must represent the variable's true maximum (which, in theory, could be infinity) such that the estimated survival function at that value is approximately zero; otherwise, data beyond that value will be cut off. Since the survival function 
is nonnegative and decreases monotonically, this cut-off can lead the existing 
conditional mean imputation approach, which we call ``non-extrapolated,'' to underestimate the integral and miscalculate the conditional means.

For example, if the last time to diagnosis was 10 years from study entry, non-extrapolated conditional mean imputation assumes that all unobserved times to diagnosis should  be observed within 10 years of study entry. Yet, in reality, diagnosis could occur at any time between the last study visit and death, both of which are unique to each subject. 
Thus, censored covariates are likely to be imputed with incorrect conditional means, leading to invalid statistical inference in the downstream analysis (e.g., when fitting a model to the imputed data). To avoid this situation, we propose several improvements to conditional mean imputation for a censored covariate.

\subsection{Need for Extrapolation with Imputation}

Many methods may come to mind that handle integrals with infinite bounds, such as Gauss--Hermite quadrature. In fact, there are many attractive methods for numerical integration already implemented in existing software that can handle infinite bounds, for example, the \texttt{integrate} function in R, which uses adaptive quadrature \citep{R}. However, even with these methods, we can only integrate over values of the covariate where the integrand is defined.


To truly improve the calculation, we need the integrand (the estimated survival function) to be defined up to the infinite bound in the conditional mean formula. Specifically, we need a way to extrapolate from Breslow's estimator beyond the largest uncensored value so that we can adopt an improved approach (in our case, adaptive quadrature) to integrate over it. Extrapolation methods are well established. However, our needs are unique: We are not just interested in extending the survival curve -- any of the ``usual'' methods like those in 
\citet{klein&moeschberger2003} could work if so -- but in further integrating over it. 

In search of the best one for our purposes, we thoroughly explored various methods to extend the survival estimator and identified the best one for our proposed ``extrapolated'' conditional mean imputation approach (Section~\ref{sims:extend}). To our knowledge, only one paper had investigated this need previously \citep{datta2005}. They considered fewer methods and, in fact, we found that their recommended method could lead to bias even when integrating up to infinity with adaptive quadrature. We note that inference about the tail of the survival function has been studied extensively \citep{Reid&Cox1984}, but we are interested in inference about the regression model after imputing based on the survival function instead. These two problems are fundamentally different. Inference about the tail requires extrapolating the survival function beyond the largest uncensored value, whereas the inference we are interested in requires extrapolating the survival function and integrating over that extrapolated function.

Importantly, extending the survival curve for integration is not a challenge unique to imputation. Any nonparametric or semiparametric full-likelihood approach with a censored covariate would also need to integrate up to infinity over an integrand that is not defined over that range. Thus, our proposed improvements hold broader implications and could be adopted to improve other methods, like a maximum likelihood estimator, as well. 

\subsection{Overview}

We propose a hybrid approach to conditional mean imputation that splices together the semiparametric survival estimator with a parametric extension, making it possible to completely approximate the integral up to infinity. 
Since the semiparametric survival estimator (Breslow's estimator) is not well defined for larger values than those in the data, we explore various extrapolation methods and identify the ``Weibull extension'' as the best one. We quantify the bias introduced by calculating conditional means from Breslow's estimator using the trapezoidal rule and show in extensive simulation studies that extrapolating from Breslow's estimator with the Weibull extension before imputation reduces bias when imputing censored covariates, even when the data were not truly Weibull. We further show how imputing with biased conditional means can impact clinical trial recruitment. The rest of the paper is as follows: we describe the proposed methods in Section~\ref{sec:methods}, we evaluate those methods against existing ones through extensive simulations in Section~\ref{sim_stud}, we apply both approaches to the analysis of Huntington's disease data from the PREDICT-HD study in Section~\ref{predict_hd}, and we discuss our findings in Section~\ref{sec:discuss}.

\section{Methods}
\label{sec:methods}

\subsection{Model and Data}\label{subsec:model&data}

Consider an outcome $Y$ and covariates ($X$, $\bZ$), which are assumed to be related through a regression model parameterized by $\btheta$ and denoted by ${\rm P}_{\btheta}(Y|X,\bZ)$. For example, if $Y$ given ($X$, $\bZ$) follows a linear regression model, 
${\rm P}_{\btheta}(Y|X,\bZ) = 1/(\sqrt{2\pi\sigma^2})\exp\{-(Y - \alpha - \beta X - \pmb{\gamma}^{\rm T}\bZ)^2/(2\sigma^2)\},$ where $\btheta = (\alpha, \beta, \pmb{\gamma}^{\rm T}, \sigma^2)^{\rm T}$. Estimating the outcome model parameters $\btheta$ is our primary interest.

Unfortunately, estimating $\btheta$ is difficult because the covariate $X$ is right-censored. Rather than observe $X$ directly, we observe $W = \min(X, C)$ and $\Delta = {\rm I}(X \leq C)$, where $C$ is a random censoring value.
(Having $C$ random rather than fixed means that $C$ changes for every subject and is unknown at study start. For Huntington's disease studies, $C$ is the subject-specific length of follow-up from first to last study visit.) Thus, an observation for subject $i$ in a sample of $n$ subjects is captured as ($Y_i, \Delta_i, W_i, \bZ_i$).

\subsection{Conditional Mean Imputation} \label{sec:imp}
In missing data settings, imputation is a popular approach to obtain valid statistical inference without sacrificing the power of the full sample. Imputation is also a promising method to handle censored covariates, with one simple change. When $W_i$ is right-censored, rather than impute any value for it, we impute a value that is larger than $W_i$ because, by the definition of right-censoring, the true unobserved $X$ must be larger than $W_i$. This partial information that $X > W_i$ is captured through a conditional mean imputation approach \citep{Little1992,Richardson&Ciampi2003}.

In conditional mean imputation, we replace right-censored covariates $W_i$ with their corresponding conditional means 
\begin{align}
{\rm E}(X|X>W_i,\bZ_i) &= W_i + \frac{\int_{W_i}^{\infty}S(x|\bZ_i){\rm d}x}{S(W_i|\bZ_i)}, \label{cm_Z} 
\end{align}
where $S(t|\bz)$ is the conditional survival function for $X = t$ given $\bZ = \bz$. To our knowledge, this form for the imputation of randomly right-censored covariates was first introduced by \citet{AtemEtAl2017}, with a thorough derivation set forth by \citet{LotspeichGrosserGarcia2022}. Previously, a parallel formula was given in \citet{Little2002} to impute covariates that are left-censored by a lower limit of detection. Note that we use the $i$ subscript for $W_i$ and $\bZ_i$ because these are observed values of random variables $W$ and $\bZ$, respectively, whereas $X$  is still random. Importantly, deriving Equation~\eqref{cm_Z} relies on the assumption of conditionally noninformative censoring, such that the censoring values $C$ and true covariates $X$ are assumed to be conditionally independent given the other fully observed covariates $\pmb{Z}$. 

Now, conditional mean imputation proceeds in two stages. First, we calculate the conditional means for all censored covariates, which requires estimating $S(t|\bz)$ (Section~\ref{sec:survival}) and approximating the integral over it (Sections~\ref{sec:pub_integral}--\ref{sec:prop_integral}). Then, we replace the censored  covariates with these conditional means and fit the outcome model for $Y$ given imputed $X$ and $\bZ$ using the ``usual'' methods (e.g., ordinary least squares) to obtain the estimators $\widehat{\btheta}$. Under proper specification (e.g., a well-estimated survival function and a well-approximated integral), \citet{Bernhardtetal2015} prove that conditional mean imputation leads to consistent estimators in linear regression (i.e., estimated $\widehat{\btheta}$ converges in probability to true $\btheta$). 

\subsection{Estimating the Survival Function}\label{sec:survival}

To robustly estimate $S(t|\bZ)$ in Equation~(\ref{cm_Z}) without assuming a distribution for $X$ given $\bZ$, and, in doing so, bypassing some potential misspecification, existing approaches use semiparametric models \citep{AtemMatsoukaZimmern2019,AtemEtAl2017,AtemSampeneGreene2019}. Specifically, existing approaches use a Cox proportional hazards model, from which the survival function can be calculated as $S(t|\bz) = S_{0}(t)^{\exp(\blambda^{\rm T}\bz)}$ with $\blambda$ the log hazard ratios and $S_{0}(t)$ the baseline survival function of $X$ (i.e., $S_0(t) \equiv S(t|\bZ = \pmb{0})$). 

This semiparametric model for $S(t|\bZ)$ requires estimating two key parts: (i) the log hazard ratios $\blambda$ and (ii) the baseline survival function $S_0(t)$. The log hazard ratios $\blambdahat$ are easily estimated from existing software, like the \texttt{coxph} function in the
\pkg{survival} package \citep{Therneau2000}, and a common way to estimate $\widehat{S}_0(t)$ is with Breslow's estimator \citep{Breslow1972}:
\begin{align}
\widehat{S}_{0}(t) &= \exp\left[- \sum_{i=1}^{n}{\rm I}(W_i \leq t)\left\{\frac{\Delta_i}{\sum_{j=1}^{n}{\rm I}(W_j \leq W_i)\exp\left(\widehat{\blambda}^{\rm T}\bZ_{j}\right)}\right\}\right]. \label{breslow}
\end{align}
After estimating $\blambdahat$ and $\widehat{S}_0(t)$, we will construct $\widehat{S}(t|\bz) = \widehat{S}_{0}(t)^{\exp(\widehat{\blambda}^{\rm T}\bz)}$ and use this estimated survival function to compute ${\rm E}(X|X>W_i,\bZ_i)$ from Equation~\eqref{cm_Z}. Still, computing this conditional mean requires a method to approximate the integral over $\widehat{S}(t|\bz)$ from $t = W_i$ to infinity.

\subsection{The Problem with Using the Trapezoidal Rule to Calculate Conditional Means}\label{sec:pub_integral}

Existing approaches use the trapezoidal rule to estimate this integral over $\widehat{S}(t|\bz)$ and compute the conditional means ${\rm E}(X|X>W_i,\bZ_i)$. That is, they estimate the integral $\int_{W_i}^{\infty}\widehat{S}_{0}(x)^{\exp(\widehat{\blambda}^{\rm T}\bZ_i)}{\rm d}x$ in Equation~\eqref{cm_Z} with
\begin{align}
\frac{1}{2}\left[\sum_{j=1}^{n-1}{\rm I}(W_{(j)}\geq W_{i})\left\{\widehat{S}_{0}(W_{(j+1)})^{\exp(\widehat{\blambda}^{\rm T}\bZ_i)} + \widehat{S}_{0}(W_{(j)})^{\exp(\widehat{\blambda}^{\rm T}\bZ_i)}\right\}\left(W_{(j+1)} - W_{(j)}\right)\right],\label{trap}
\end{align}
where $W_{(1)} < \cdots < W_{(n)}$ denote the $n$ distinct, ordered values of $W$ from the data. Going forward, let the conditional mean following the trapezoidal rule be $\widehat{\rm E}(X|X>W_i,\bZ_i)=$
\begin{align*}
& W_i + \frac{1}{2}\left(\frac{\left[\sum_{j=1}^{n-1}{\rm I}(W_{(j)}\geq W_{i})\left\{\widehat{S}_{0}(W_{(j+1)})^{\exp(\widehat{\blambda}^{\rm T}\bZ_i)} + \widehat{S}_{0}(W_{(j)})^{\exp(\widehat{\blambda}^{\rm T}\bZ_i)}\right\}\left(W_{(j+1)} - W_{(j)}\right)\right]}{\widehat{S}_{0}(W_{i})^{\exp(\widehat{\blambda}^{\rm T}\bZ_i)}}\right).
\end{align*}
This formula for the conditional mean is prominent in the current literature around imputing randomly right-censored covariates \citep{AtemMatsoukaZimmern2019,AtemEtAl2017,AtemSampeneGreene2019,LotspeichGrosserGarcia2022}. Herein, we refer to these existing approaches involving imputation with $\widehat{\rm E}(X|X>W_i,\bZ_i)$ as non-extrapolated conditional mean imputation. 

Notice that the ``trapezoids'' in Expression~\eqref{trap} are defined between the observed values $W_{(j)} \geq W_{i}$ and their estimated survival functions given the $i$th subject's covariates, $\widehat{S}(W_{(j)}|\bZ_i)$. Some $W_{(j)}$ will be censored, so computing $\widehat{\rm E}(X|X>W_i,\bZ_i)$ requires evaluating $\widehat{S}_0(\cdot)$ between and beyond the uncensored data on which it is defined. Between uncensored values, $\widehat{S}_0(\cdot)$ should be carried forward (interpolated) from the last uncensored value. Beyond the largest uncensored value, $\widehat{S}_0(\cdot)$ is defined to carry forward, but that can be unrealistic; we consider multiple methods to extrapolate from it in Section~\ref{sec:prop_integral}. \bigskip 

\noindent\textit{Remark} 2.1.
Instead of using Breslow's estimator as defined, the existing approaches (e.g., \citet{AtemMatsoukaZimmern2019}) interpolate with the mean of $\widehat{S}_0(\cdot)$ from the uncensored values immediately below and above a censored $W_{(j)}$. Here, we will adopt carry forward interpolation because it is computationally simple and follows from the original formula in \citet{Breslow1972}, although we show in Section~\ref{sims:extend} that either mean or carry forward interpolation seems to work well.

\bigskip 

Critically, we recognize that this use of the trapezoidal rule in Expression~\eqref{trap} estimates the wrong integral: $\int_{W_i}^{W_{(n)}}\widehat{S}_{0}(x)^{\exp(\widehat{\blambda}^{\rm T}\bZ_i)}{\rm d}x$ rather than the targeted $\int_{W_i}^{\infty}\widehat{S}_{0}(x)^{\exp(\widehat{\blambda}^{\rm T}\bZ_i)}{\rm d}x$. The validity of this estimate, and with it the quality of the conditional means, hinges on how well the maximum of the observed covariate $W_{(n)}$ represents the true maximum of the covariate $X$; this sentiment is shared in \citet{AtemEtAl2017}. If $W_{(n)}$ is far below the true upper bound of $X$, then approximating with $\int_{W_i}^{W_{(n)}}\widehat{S}_{0}(x)^{\exp(\widehat{\blambda}^{\rm T}\bZ_i)}{\rm d}x$ will underestimate the integral by cutting off the tail of the survival function. We conclude that using the trapezoidal rule to calculate conditional means is only appropriate when $\widehat{S}_{0}(W_{(n)}) \approx 0$, because in this case the survival function is entirely captured by $W_{(1)} < \cdots < W_{(n)}$. 
Therefore, we set out to propose a more general approach to correctly calculate conditional means even when $\widehat{S}_{0}(W_{(n)}) > 0$.

\subsection{Replacing the Trapezoidal Rule with Adaptive Quadrature}\label{sec:prop_integral}

We sought an improved calculation to capture the entirety of the improper integral in the conditional means by extending beyond $W_{(n)}$ to better approximate the infinite upper bound. Conveniently, the \texttt{integrate} function in R implements ``adaptive quadrature of functions ... over a finite or infinite interval'' \citep{Piessens1983,R}. This function is included in the basic R functions and does not require installing any additional packages, making it an accessible and sustainable software choice. Telling the \texttt{integrate} function that we want an infinite upper bound is simple enough. In fact, as a user, it is no different than with a finite one. 

Still, adopting software that can integrate up to infinity does us no good if the integrand, i.e., the survival function of the censored covariate, is not defined as such; this is a problem not just for \texttt{integrate} but for all quadrature software. Before using adaptive quadrature with an infinite upper bound, we have to ``extend'' (i.e., extrapolate from) Breslow's estimator beyond the largest uncensored covariate value $\widetilde{X} = \max(W_1\Delta_1,\dots,W_n\Delta_n)$. This way, we will give the \texttt{integrate} function something to integrate over on its way up to infinity and better calculate the conditional means, as desired.

\subsection{Extending Breslow's estimator beyond the largest uncensored value}\label{sec:extrap}

We sought a method to extend Breslow's estimator beyond the largest uncensored covariate value $\widetilde{X}$, i.e., to extrapolate from $\widehat{S}_0(t)$ for values of $t$ up to infinity. Extrapolating from step functions is a common challenge with censored outcomes, since popular estimators, like Kaplan--Meier, are not well defined for values of $t > \widetilde{X}$, either \citep{klein&moeschberger2003}. We discuss four potential methods to extend Breslow's estimator.
\begin{itemize}
    \item[]\textit{Carry forward}: Carry forward Breslow's estimator from $\widetilde{X}$. By estimating $\widehat{S}_{0}(t) = \widehat{S}_{0}\left(\widetilde{X}\right)$ for all $t > \widetilde{X}$, this asserts that all censored covariates would have had $X = \infty$. 
    \item[]\textit{Immediate drop-off}: Do not extrapolate from Breslow's estimator at all. Assuming that $\widehat{S}_{0}(t) = 0$ at all $t > \widetilde{X}$ is equivalent to assuming that the true values $X$ for all censored covariates would have fallen just beyond their observed values $W_i$. 
    \item[]\textit{Exponential extension}: ``Tie in'' an exponential survival function where Breslow's estimator leaves off and assume that $\widehat{S}_{0}(t) = \exp\left(\left[t \log\left\{\widehat{S}_{0}(\widetilde{X})\right\}\right]/\widetilde{X}\right)$ for $t > \widetilde{X}$. 
    \item[]\textit{Weibull extension}: For added flexibility, tie in a Weibull survival function 
    and assume that $\widehat{S}_{0}(t) = \exp\left(-\hat{\rho} t^{\hat{\nu}}\right)$ for $t > \widetilde{X}$, where $\hat{\nu}$ and $\hat{\rho}$ are found using constrained maximum likelihood estimation \citep{Moeschberger&Klein1985}. 
\end{itemize}
While these methods are well established for censored outcomes, to our knowledge we are the first to consider them for censored covariates. Also, our needs are unique, since we are extrapolating from the survival curve to then integrate over it. 
Without an extrapolation method, improving the conditional mean calculation from a step survival function like Breslow's estimator would be impossible; no matter how well we can integrate up to infinity, the integrand must be defined across the entire range, which requires extrapolation.

Either carry forward or immediate drop-off could be a valid modification if we were just modeling the survival function, since they can converge to the true survival functions in large samples \citep{ying1989,klein&moeschberger2003}. However, neither is a good choice when we are subsequently integrating over the survival function. Carry forward makes the integral up to infinity diverge. Immediate drop-off forces the integral to cut off at $\widetilde{X}$; therefore, we expect it to offer little improvement over the trapezoidal rule, even with adaptive quadrature. (This is the method recommended by \citet{datta2005} for integration under the Kaplan--Meier estimator, and we show empirically in Section~\ref{sims:extend} that our expectation of its performance held true.) Fortunately, theoretical justification exists for both parametric extensions, so we explored them in extensive simulations before making recommendations (Section~\ref{sims:extend}). Derivations for the parametric extensions can be found in Web Appendix A, along with an illustration of these extrapolation methods (Supplemental Figure~S1).

\bigskip 
\noindent \textit{Remark} 2.2. 
Calculating conditional means with the trapezoidal rule can still involve evaluating the survival function for values of $t > \widetilde{X}$. In the absence of additional covariates $\bZ$, the existing approaches (e.g., \citet{AtemMatsoukaZimmern2019}) treat the largest value $W_{(n)}$ as uncensored regardless of $\Delta_{(n)}$, a recommendation from \citet{datta2005}, so that the Kaplan--Meier estimator equals zero at $W_{(n)}$. 
This method is equivalent to immediate drop-off but its impact is subtle, since the trapezoidal rule cuts the tail off anyway. To our knowledge, the existing approaches do not define an extrapolation method for Breslow's estimator when covariates $\bZ$ are available. 

\section{Simulation Studies}\label{sim_stud}

Before we can use adaptive quadrature with an infinite upper bound (hereafter called ``adaptive quadrature''), we must decide how to extrapolate from Breslow's estimator. We choose the Weibull extension, which we show offers low bias and high efficiency in the downstream analysis even when $X$ given $\bZ$ is not truly Weibull (Section~\ref{sims:extend}). Then, we highlight the improvements (i.e., substantially reduced bias and some heightened efficiency) of extrapolated versus non-extrapolated conditional mean imputation 
(Section~\ref{sims:compare_approaches}). R scripts to reproduce all simulations, tables, and figures, along with all simulated data, are available on GitHub at \underline{https://github.com/sarahlotspeich/hybridCMI}.

\subsection{Data Generation and Metrics for Comparison}\label{data_gen}

We simulated data for samples of $n = 100$, $500$, $1000$, or $2000$ subjects in the following way. First, a binary covariate $Z$ was generated from a Bernoulli distribution with ${\rm P}(Z=1) = 0.5$. Next, $X$ was generated from a Weibull distribution with shape $= 0.75$ and scale $= 0.25 + 0.25Z$, leading to proportional hazards in $X$ given $Z$. Then, a continuous outcome was generated as $Y = 1 + 0.5X + 0.25Z + e$, where $e$ was a standard normal random variable. We explored light ($\sim 17\%$), heavy ($\sim 49\%$), and extra heavy ($\sim 82\%$) censoring in $X$, induced by generating $C$ from an exponential distribution with rates $= 0.5$, $2.9$, and $20$, respectively. See Supplemental Figure~S2 for a summary of censoring rates across simulations. Notice that $C$ was generated independently of all other variables, which more than satisfies our assumption of conditionally noninformative censoring. Finally, $W = \min(X, C)$ and $\Delta = {\rm I}(X \leq C)$ were constructed. 

Given a continuous outcome $Y$, the analysis model ${\rm P}_{\btheta}(Y|X,\bZ)$ was a linear regression. We considered two imputation approaches to estimate $\widehat{\btheta}$: one using the extrapolated survival curve and adaptive quadrature, called extrapolated conditional mean imputation, and the other using the non-extrapolated survival function and the trapezoidal rule, called non-extrapolated conditional mean imputation. To assess validity, we report the empirical bias and standard errors for both $\bthetahat$. To gauge statistical precision, we report the relative efficiency, which was calculated as the empirical variance of the full cohort analysis (i.e., where all $n$ observations had uncensored $X$) divided by the empirical variance of the imputation approaches. The closer the relative efficiency is to one, the more efficiency was recovered through imputation. Unless otherwise stated, all summary metrics (bias, standard errors, and relative efficiency) are based on \num{1000} replications.

Our simulation settings are based on those of \citet{AtemEtAl2017}, who, to the best of our knowledge were the first to propose (non-extrapolated) conditional mean imputation for a randomly right-censored covariate in a linear regression like ours. However, there are a few distinctions to note. First, $X$ was generated conditionally on $Z$, whereas \citet{AtemEtAl2017} generated $X$ from either (i) a Weibull distribution with constant shape and scale (independently of $Z$) or (ii) a Weibull distribution with shape dependent on $Z$ and constant scale (leading to non-proportional hazards in $X$ given $Z$). Second, an additional set of simulations under ``extra heavy'' censoring were considered here, chosen to reflect the severe censoring rate in the PREDICT-HD dataset.

\subsection{Extending the Estimated Survival Function: How to Extrapolate from Breslow's Estimator}\label{sims:extend}

To extend Breslow's estimator, we considered three of the extrapolation methods for $\widehat{S}_0(t)$ introduced in Section~\ref{sec:extrap}: (i) immediate drop-off, (ii) exponential extension, and (iii) Weibull extension. (We did not consider carry forward extrapolation, since it caused the integral to diverge.) To compare them, we focused on estimating $\beta$, the coefficient on $X$, which will be most impacted by censoring. Extrapolating $\widehat{S}_0(t)$ with the Weibull extension offered the lowest bias and best efficiency for the extrapolated conditional mean imputation estimator $\hat{\beta}$ (Supplemental Figure~S3).

Though the ``winning'' method used the Weibull extension to extrapolate, $X$ was truly generated from a Weibull distribution here. Therefore, to offer more general recommendations, we also considered an $X$ that was generated from a log-normal distribution with mean $= 0.05Z$ and variance $= 0.25$ (on the log scale). For light ($\sim 20\%$), moderate ($\sim 35\%$), and heavy ($\sim 80\%$) censoring, we generated $C$ from an exponential distribution with rates $= 0.2$, $0.4$, and $1.67$, respectively. The parameters used to generate log-normal $X$ were chosen to achieve similar censoring rates with Weibull $X$ in the light, heavy, and extra heavy settings. Interestingly, with log-normal $X$, the bias when using extrapolated conditional mean imputation was very low and relatively unchanged by the extrapolation methods (Supplemental Figure~S4). 

For another example where the extrapolation approach was misspecified for the data generating mechanism, see Supplemental Figure~S3 where the exponential extension still offered reduced bias over immediate drop-off even when $X$ was truly Weibull. In this example, the parametric extension of the survival curve assumed constant hazard for $X$ across $Z$, whereas $X$ was simulated to have proportional hazards across $Z$. Still, extrapolated conditional mean imputation offered reduced bias over non-extrapolated conditional mean imputation. 

We also compared mean versus carry forward interpolation between uncensored values for Breslow's estimator (Remark~2.1) and found that they performed similarly in terms of bias and efficiency in $\hat{\beta}$ (Supplemental Figure~S5). Also, as expected in Remark~2.2, there were only minor differences between the extrapolation methods when using non-extrapolated conditional mean imputation (Supplemental Figure~S6). Now, armed with the Weibull extension, we can extend Breslow's estimator to infinity and proceed with comparing our proposed extrapolated conditional mean imputation approach to the existing non-extrapolated approach in a variety of real-world scenarios.

\subsection{Quantifying the Improvement: Extrapolated Versus Non-Extrapolated Conditional Mean Imputation} \label{sims:compare_approaches}

Having selected the Weibull extension method for extrapolation, we compared the resulting linear regression estimates between extrapolated and non-extrapolated conditional mean imputation approaches. 
After estimating the survival function for Weibull $X$, non-extrapolated conditional mean imputation 
led to large bias in $\hat{\beta}$ 
(Table~\ref{table:weibX}). Under light, heavy, and extra heavy censoring, this approach led to as much as 20\%, 25\%, and 200\% bias, 
respectively. Meanwhile, extrapolated conditional mean imputation 
offered no more than 4\%, 18\%, and 44\% bias under light, heavy, and extra heavy censoring, respectively. 
With minor exceptions (e.g., in the largest samples), extrapolated conditional mean imputation 
continued to have efficiency gains over non-extrapolated conditional mean imputation 
even when estimating $S(t|z)$. When $X$ was generated independently of $Z$, non-extrapolated conditional mean imputation could lead to unbiased estimates for $\alpha$ and $\gamma$ but continued to see bias (as high as 149\%) in estimating $\beta$ (Supplemental Table~S1). 

\begin{table}
\caption{Simulation results for Weibull $X$ from the full cohort analysis and imputation approaches. 
\label{table:weibX}}
\centering
\resizebox{\columnwidth}{!}{
\begin{threeparttable}
\begin{tabular}{lrlrrclrrcclrrcc}
\hline 
& & & \multicolumn{3}{c}{} && \multicolumn{4}{c}{\textbf{Extrapolated Conditional}} && \multicolumn{4}{c}{\textbf{Non-Extrapolated Conditional}} \\
& & & \multicolumn{3}{c}{\textbf{Full Cohort}} && \multicolumn{4}{c}{\textbf{ Mean Imputation}} && \multicolumn{4}{c}{\textbf{ Mean Imputation}} \\
\cmidrule(l{3pt}r{3pt}){4-6} \cmidrule(l{3pt}r{3pt}){8-11} \cmidrule(l{3pt}r{3pt}){13-16}
\textbf{Censoring} & $\pmb{n}$ && \textbf{Bias} & \textbf{(\%)} & \textbf{SE} && \textbf{Bias} & \textbf{(\%)} & \textbf{SE} & \textbf{RE} && \textbf{Bias} & \textbf{(\%)} &  \textbf{SE} & \textbf{RE}\\ \hline 
\multicolumn{16}{c}{\textbf{$\pmb{\hat{\alpha}}$: Intercept}} \\
\addlinespace
Light & 100 &  & $0.000$ & ($0.01$) & $0.153$ &  & $0.003$ & ($0.30$) & $0.157$ & $0.951$ &  & $-0.021$ & ($-2.12$) & $0.162$ & $0.897$\\
& 500 &  & $-0.002$ & ($-0.25$) & $0.065$ &  & $0.003$ & ($0.32$) & $0.067$ & $0.918$ &  & $-0.014$ & ($-1.38$) & $0.069$ & $0.878$\\
& 1000 &  & $0.001$ & ($0.05$) & $0.048$ &  & $0.006$ & ($0.58$) & $0.049$ & $0.947$ &  & $-0.005$ & ($-0.52$) & $0.051$ & $0.859$\\
& 2000 &  & $0.001$ & ($0.07$) & $0.034$ &  & $0.005$ & ($0.48$) & $0.036$ & $0.915$ &  & $-0.002$ & ($-0.22$) & $0.037$ & $0.834$\\
Heavy & 100 &  & $0.000$ & ($0.01$) & $0.153$ &  & $-0.019$ & ($-1.91$) & $0.174$ & $0.773$ &  & $-0.056$ & ($-5.59$) & $0.189$ & $0.657$\\
& 500 &  & $-0.002$ & ($-0.25$) & $0.065$ &  & $-0.015$ & ($-1.54$) & $0.074$ & $0.761$ &  & $-0.053$ & ($-5.34$) & $0.079$ & $0.666$\\
& 1000 &  & $0.001$ & ($0.05$) & $0.048$ &  & $-0.008$ & ($-0.83$) & $0.054$ & $0.760$ &  & $-0.046$ & ($-4.56$) & $0.057$ & $0.695$\\
& 2000 &  & $0.001$ & ($0.07$) & $0.034$ &  & $-0.005$ & ($-0.45$) & $0.039$ & $0.764$ &  & $-0.042$ & ($-4.22$) & $0.041$ & $0.680$\\
Extra Heavy & 100 &  & $0.000$ & ($0.01$) & $0.153$ &  & $0.040$ & ($3.96$) & $0.209$ & $0.537$ &  & $-0.065$ & ($-6.46$) & $0.284$ & $0.290$\\
& 500 &  & $-0.002$ & ($-0.25$) & $0.065$ &  & $0.007$ & ($0.69$) & $0.094$ & $0.475$ &  & $-0.080$ & ($-8.01$) & $0.119$ & $0.294$\\
& 1000 &  & $0.001$ & ($0.05$) & $0.048$ &  & $-0.002$ & ($-0.20$) & $0.073$ & $0.428$ &  & $-0.076$ & ($-7.60$) & $0.086$ & $0.306$\\
& 2000 &  & $0.001$ & ($0.07$) & $0.034$ &  & $-0.006$ & ($-0.56$) & $0.053$ & $0.412$ &  & $-0.069$ & ($-6.88$) & $0.063$ & $0.297$\\
\addlinespace
\multicolumn{16}{c}{\textbf{$\pmb{\hat{\beta}}$: Coefficient on Censored $\pmb{X}$}} \\
\addlinespace
Light & 100 &  & $-0.003$ & ($-0.54$) & $0.177$ &  & $-0.020$ & ($-3.94$) & $0.216$ & $0.673$ &  & $-0.019$ & ($-3.88$) & $0.224$ & $0.628$\\
& 500 &  & $0.001$ & ($0.26$) & $0.070$ &  & $-0.021$ & ($-4.20$) & $0.095$ & $0.542$ &  & $-0.069$ & ($-13.87$) & $0.099$ & $0.507$\\
& 1000 &  & $0.003$ & ($0.64$) & $0.050$ &  & $-0.017$ & ($-3.36$) & $0.065$ & $0.584$ &  & $-0.090$ & ($-17.96$) & $0.075$ & $0.440$\\
& 2000 &  & $0.000$ & ($-0.08$) & $0.036$ &  & $-0.016$ & ($-3.17$) & $0.051$ & $0.496$ &  & $-0.102$ & ($-20.47$) & $0.059$ & $0.372$\\
Heavy & 100 &  & $-0.003$ & ($-0.54$) & $0.177$ &  & $0.091$ & ($18.24$) & $0.373$ & $0.226$ &  & $0.127$ & ($25.46$) & $0.417$ & $0.180$\\
& 500 &  & $0.001$ & ($0.26$) & $0.070$ &  & $0.061$ & ($12.29$) & $0.160$ & $0.192$ &  & $0.038$ & ($7.58$) & $0.159$ & $0.194$\\
& 1000 &  & $0.003$ & ($0.64$) & $0.050$ &  & $0.042$ & ($8.39$) & $0.121$ & $0.170$ &  & $0.000$ & ($0.06$) & $0.112$ & $0.198$\\
& 2000 &  & $0.000$ & ($-0.08$) & $0.036$ &  & $0.020$ & ($3.99$) & $0.086$ & $0.174$ &  & $-0.028$ & ($-5.53$) & $0.080$ & $0.201$\\
Extra Heavy & 100 &  & $-0.003$ & ($-0.54$) & $0.177$ &  & $-0.219$ & ($-43.77$) & $0.650$ & $0.074$ &  & $1.001$ & ($200.27$) & $1.875$ & $0.009$\\
& 500 &  & $0.001$ & ($0.26$) & $0.070$ &  & $-0.023$ & ($-4.52$) & $0.358$ & $0.038$ &  & $0.784$ & ($156.79$) & $0.604$ & $0.013$\\
& 1000 &  & $0.003$ & ($0.64$) & $0.050$ &  & $0.076$ & ($15.29$) & $0.325$ & $0.024$ &  & $0.690$ & ($137.93$) & $0.421$ & $0.014$\\
& 2000 &  & $0.000$ & ($-0.08$) & $0.036$ &  & $0.117$ & ($23.41$) & $0.267$ & $0.018$ &  & $0.583$ & ($116.66$) & $0.305$ & $0.014$\\
\addlinespace
\multicolumn{16}{c}{\textbf{$\pmb{\hat{\gamma}}$: Coefficient on Uncensored $\pmb{Z}$}} \\
\addlinespace
Light & 100 &  & $0.001$ & ($0.33$) & $0.210$ &  & $0.003$ & ($1.18$) & $0.214$ & $0.962$ &  & $0.062$ & ($24.95$) & $0.210$ & $0.993$\\
& 500 &  & $0.002$ & ($0.73$) & $0.092$ &  & $0.002$ & ($0.72$) & $0.094$ & $0.955$ &  & $0.074$ & ($29.58$) & $0.094$ & $0.942$\\
& 1000 &  & $-0.003$ & ($-1.25$) & $0.064$ &  & $-0.003$ & ($-1.25$) & $0.065$ & $0.973$ &  & $0.074$ & ($29.49$) & $0.066$ & $0.937$\\
& 2000 &  & $-0.001$ & ($-0.30$) & $0.047$ &  & $0.000$ & ($-0.11$) & $0.047$ & $0.960$ &  & $0.077$ & ($30.80$) & $0.048$ & $0.945$\\
Heavy & 100 &  & $0.001$ & ($0.33$) & $0.210$ &  & $0.018$ & ($7.28$) & $0.223$ & $0.887$ &  & $0.118$ & ($47.22$) & $0.211$ & $0.986$\\
& 500 &  & $0.002$ & ($0.73$) & $0.092$ &  & $0.011$ & ($4.55$) & $0.100$ & $0.847$ &  & $0.124$ & ($49.79$) & $0.094$ & $0.951$\\
& 1000 &  & $-0.003$ & ($-1.25$) & $0.064$ &  & $0.003$ & ($1.14$) & $0.068$ & $0.877$ &  & $0.122$ & ($48.92$) & $0.066$ & $0.940$\\
& 2000 &  & $-0.001$ & ($-0.30$) & $0.047$ &  & $0.003$ & ($1.27$) & $0.051$ & $0.828$ &  & $0.125$ & ($49.83$) & $0.047$ & $0.981$\\
Extra Heavy & 100 &  & $0.001$ & ($0.33$) & $0.210$ &  & $-0.024$ & ($-9.76$) & $0.346$ & $0.368$ &  & $0.142$ & ($56.91$) & $0.212$ & $0.977$\\
& 500 &  & $0.002$ & ($0.73$) & $0.092$ &  & $-0.012$ & ($-4.99$) & $0.130$ & $0.495$ &  & $0.146$ & ($58.41$) & $0.094$ & $0.952$\\
& 1000 &  & $-0.003$ & ($-1.25$) & $0.064$ &  & $-0.010$ & ($-3.84$) & $0.092$ & $0.487$ &  & $0.143$ & ($57.23$) & $0.066$ & $0.931$\\
& 2000 &  & $-0.001$ & ($-0.30$) & $0.047$ &  & $0.006$ & ($2.48$) & $0.065$ & $0.508$ &  & $0.144$ & ($57.60$) & $0.047$ & $0.979$\\
\bottomrule
\end{tabular}
\begin{tablenotes}[flushleft]
\item{\em Note:} \textbf{Bias (\%)}: empirical bias (empirical percent bias); \textbf{SE}: empirical standard error; \textbf{RE}: empirical relative efficiency to the full-cohort analysis. True parameter values were $(\alpha, \beta, \gamma) = (1, 0.5, 0.25)$. The MLE for the Weibull extension converged in $\geq 99.4\%$ of replicates of imputation in each setting (just \num{24} of \num{12000} total replicates did not converge); all other entries are based on 1000 replicates.
\end{tablenotes}
\end{threeparttable}
}
\end{table}

Estimating the survival function for log-normal $X$ led to less bias for both imputation approaches (Supplemental Table~S2). Although, non-extrapolated conditional mean imputation remained more biased than extrapolated conditional mean imputation, with bias up to 16\% versus 4\% for $\beta$. We were surprised, as we expected non-extrapolated conditional mean imputation to continue to produce high bias; upon further investigation, we discovered that this was due to the symmetry of the log-normal distribution and the data generating mechanism for the censoring variable $C$. Due to the Weibull distribution's skewness, higher censoring rates, driven by larger rate parameters for $C$, led to smaller values of $W_{(n)}$ (the maximum of the observed covariate), which led to worse performance (i.e., higher bias) when calculating the conditional mean with the trapezoidal rule up to this value for non-extrapolated conditional mean imputation (Supplemental Figure~S7). Due to symmetry, the log-normal data generation continued to lead to larger values of $W_{(n)}$, even under heavy or extra heavy censoring, which could explain the improvements to with extrapolation (because less extrapolation is needed) and without (because less of the tail is cut off). 


\section{Application to Huntington's Disease Data}\label{predict_hd}

\subsection{Designing Clinical Trials to Test Experimental Treatments}

Damage due to Huntington's disease is irreversible, so slowing symptom progression is often the objective of experimental treatments. Clinical trials are critical to the success of potential treatments but also expensive, leading to constraints in their design and implementation, like the number of subjects recruited and length of follow-up. Thus, clinical trials seek to recruit subjects for whom the treatment could have the greatest potential impact \citep{Paulsen2019}.

Recruiting from an existing Huntington's disease study can be a powerful first step. 
For example, we could measure symptom change leading up to potential recruitment. Information about symptom change is important, since the impact of the treatment in slowing symptom progression would be more measurable for subjects with steeply progressing symptoms. Still, an existing study only tells us how a subject's symptoms have been changing thus far, while what we really want to know is how their symptoms would change during the trial. While this future symptom progression is not measurable, it is estimable. Specifically, we can model between-visit symptom change using data from PREDICT-HD. Then, we can use that model to estimate subjects' post-recruitment symptom progression and identify high priority subjects for a new clinical trial (i.e., those with the largest expected declines).

Time to diagnosis has been shown to be highly predictive of symptom severity, with the steepest change in symptoms seen in the years immediately before and after diagnosis (e.g., \citet{long2014tracking}). Thus, time to diagnosis is an important covariate in our symptom progression model, but in a prospective study like PREDICT-HD, where not everyone has been diagnosed, it is a randomly right-censored covariate that must first be dealt with. In the sections that follow, we discuss the details of modeling the progression of Huntington's disease symptoms in a prospective study of diagnosed and undiagnosed subjects using data from PREDICT-HD (Section~\ref{data:model}). Then, we walk through imputing censored times to diagnosis for undiagnosed subjects (Section~\ref{data:impute}). Finally, we discuss our strategy to recruit subjects for a new clinical trial based on these models (Section~\ref{data:recruit}).

\subsection{Modeling the Progression of Huntington's Disease Symptoms}\label{data:model}

One way to gauge symptom severity is the composite Unified Huntington Disease Rating Scale (cUHDRS), which collectively measures functional, motor, and cognitive impairments. As Huntington's disease progresses toward diagnosis, impairment worsens and the cUHDRS is designed to decrease as it does. Following from \citet{schobel2017motor}, $\texttt{cUHDRS} = {(\tt TFC} - 10.4)/1.9 - (\texttt{TMS} - 29.7)/14.9 + (\texttt{SDMT} - 28.4)/11.3 + (\texttt{SWR} - 66.1)/20.1 + 10$, where \texttt{TFC} is total functional capacity, \texttt{TMS} is total motor score, \texttt{SDMT} is the Symbol Digit Modality Test, and \texttt{SWR} is the Stroop Word Reading Test. These components measure symptom severity in different areas of life: capacity for ``everyday tasks'' (\texttt{TFC}), motor impairment (\texttt{TMS}), and cognitive impairment (\texttt{SDMT} and \texttt{SWR}).

We captured Huntington's disease symptom progression by modeling the adjusted association between a subject's cUHDRS at two time points (denoted by \texttt{cUHDRS\_start} and \texttt{cUHDRS\_end}), controlling for other known covariates. To fit this model, subjects' cUHDRS scores at their first and last PREDICT-HD study visits were taken as \texttt{cUHDRS\_start} and \texttt{cUHDRS\_end}, respectively. The additional covariates in this model were (i) proximity to diagnosis, defined as \texttt{TIME\_end} from the end time point to diagnosis, and (ii) baseline information about age, CAG repeat length, and their interaction (denoted by \texttt{AGE}, \texttt{CAG}, and \texttt{AGE}$\times$\texttt{CAG}, respectively). Age and CAG repeat length were both measured at first study visit. In addition, we included an interaction between \texttt{cUHDRS\_start} and \texttt{TIME\_end}. This interaction allows the cUHDRS of a subject who is farther from diagnosis to not change much, while the cUHDRS of a subject who is closer to diagnosis can change noticeably. Thus, the symptom progression model of interest was captured with linear regression as 
\begin{align}
& {\rm E}_{\btheta}(\texttt{cUHDRS\_end}|\texttt{TIME\_end}, \texttt{cUHDRS\_start}, \texttt{AGE}, \texttt{CAG}) \nonumber \\
&= \alpha + \beta \texttt{TIME\_end} + \gamma_0 \texttt{cUHDRS\_start} + \gamma_1 \texttt{TIME\_end} \times \texttt{cUHDRS\_start}  \nonumber \\
&\phantom{=} + \gamma_2 \texttt{AGE} + \gamma_3 \texttt{CAG} + \gamma_4 \texttt{AGE}\times\texttt{CAG}.\label{hd_model}
\end{align}
Covariates were rescaled to make the model intercept $\alpha$ more interpretable, with \texttt{AGE}, \texttt{CAG}, and \texttt{cUHDRS\_start} centered at $18$, $36$, and $23.8$, respectively. 

To be included in our analysis, subjects needed to have (i) a \texttt{CAG} repeat length $\geq 36$ on the HTT gene, (ii) not yet been diagnosed with Huntington's disease at study entry, (iii) undergone all necessary testing to calculate the cUHDRS at the first and last visits (Supplemental Figure~S8), and (iv) returned for at least one follow-up visit. These criteria left a sample of $n = 970$ at-risk subjects, $238$ (25\%) of whom were diagnosed before their last visit, leaving 75\% with a censored time to diagnosis covariate. Since we employed single conditional mean imputation to replace censored times to diagnosis, we estimated the robust sandwich variance with the \pkg{sandwich} package \citep{Zeileis2004}. 

\subsection{Imputing Censored Times to Diagnosis}\label{data:impute}

\begin{figure}
    \includegraphics[width=0.95\textwidth]{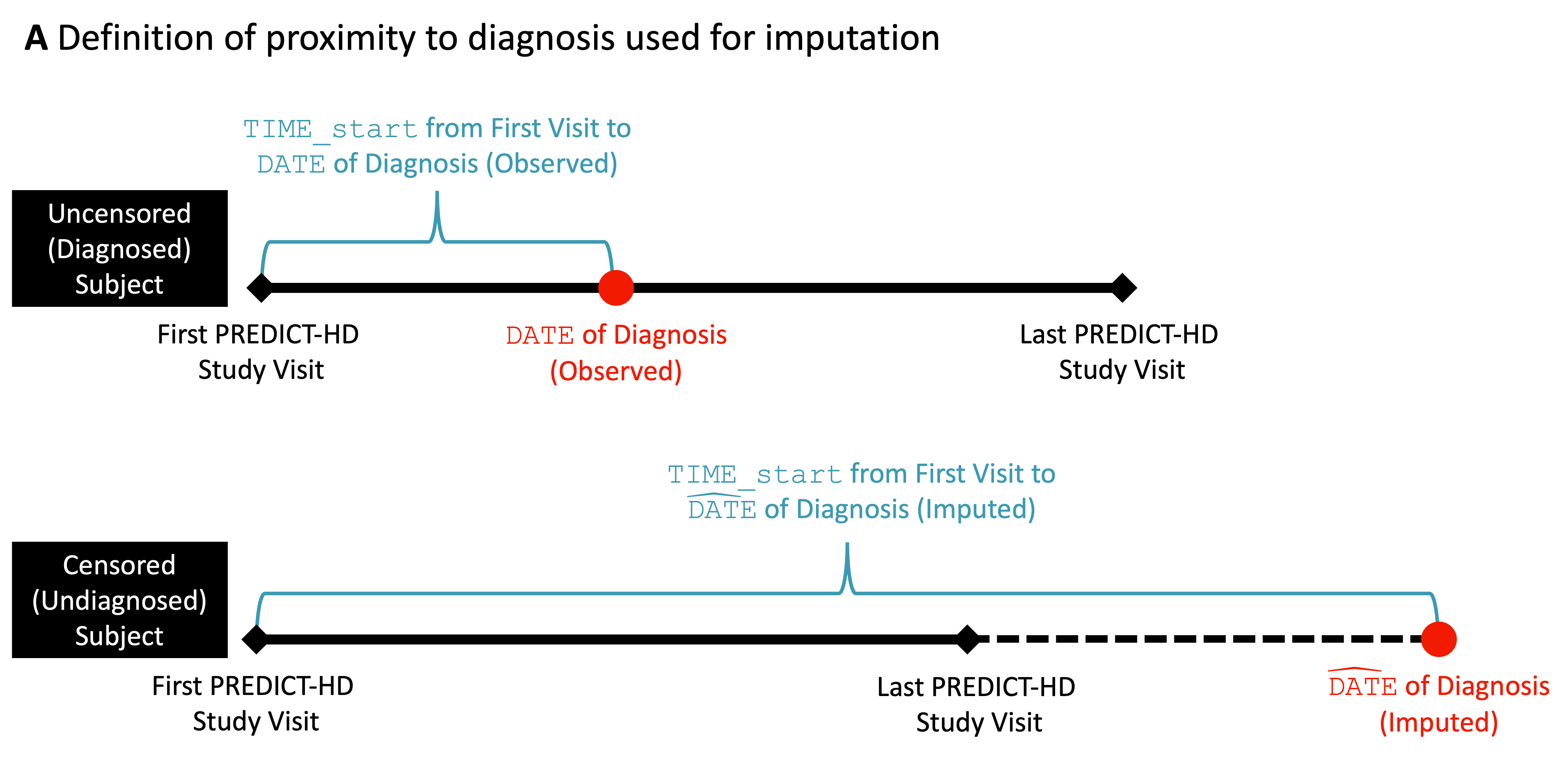}
    \includegraphics[width=0.95\textwidth]{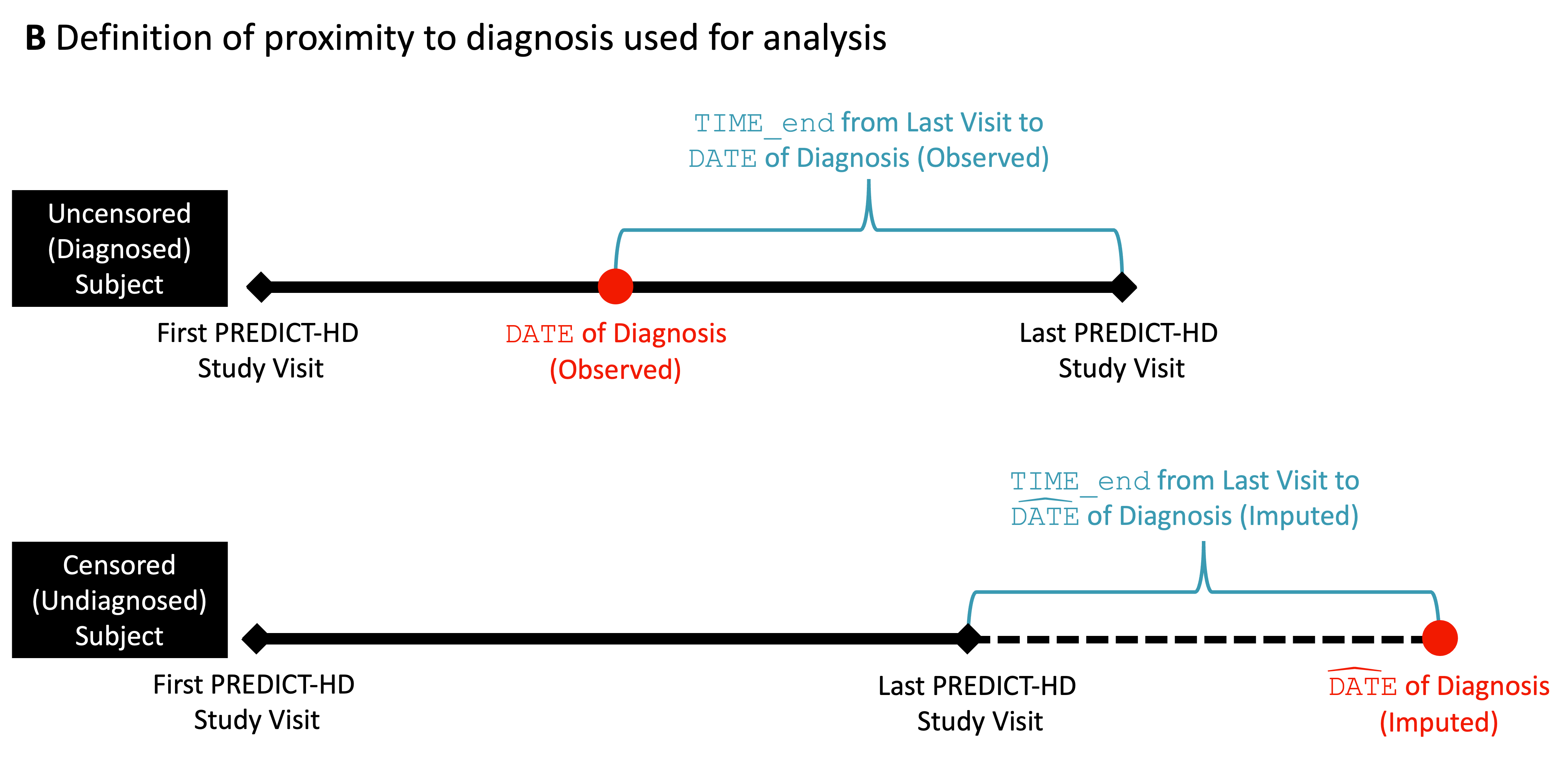}
    \caption{From a subject's observed or imputed date of diagnosis, we calculated their time to diagnosis from either the first (\textbf{A}) or last visit (\textbf{B}).}
    \label{fig:define-time-to-diagnosis}
\end{figure}

Calculating time to diagnosis was done in the following way. First, \texttt{DATE} of diagnosis was taken as the first visit where a subject met the criteria for diagnosis, i.e., a clinician assigned them to the highest rating of a $4$ on the Unified Huntington's Disease Rating Scale diagnostic confidence level \citep{long2014tracking}. From \texttt{DATE}, we calculated time to diagnosis from either the start or end of the time period, denoted as \texttt{TIME\_start} and \texttt{TIME\_end}, respectively (Figure~\ref{fig:define-time-to-diagnosis}). We did the former for imputation, because it was most natural to think of the symptom progression from the start of the period. We did the latter for analysis, because time from the end of the period aligned better with our outcome (cUHDRS at that same time).

Since subjects who had not yet been diagnosed had no such \texttt{DATE} but would have one someday in the future, \texttt{TIME\_start} from the start of the period to diagnosis was randomly right-censored. This variable was imputed for undiagnosed subjects with their conditional means ${\rm E}(\texttt{TIME\_start}|\texttt{TIME\_start} > \texttt{FOLLOW\_UP}, \texttt{AGE}, \texttt{CAG})$, where $\texttt{FOLLOW\_UP}$ was their disease-free follow-up time from the start to the end of the period. Imputation began by modeling the conditional survival function for \texttt{TIME\_start} given other fully observed covariates (\texttt{AGE}, \texttt{CAG}) from study entry. First, we fit the Cox proportional hazards model and calculated Breslow's estimator (details in Web Appendix C.1). Following from our empirical findings in Section~\ref{sims:extend}, we used the Weibull extension to extrapolate the survival estimator beyond the largest uncensored value, where $\widehat{S}_0(t=11.42) = 0.89$. Also, the context of \texttt{TIME\_start} could be used to refine the upper bound of the integral. 
Specifically, \texttt{TIME\_start} from start of the time period to Huntington's disease diagnosis could not be infinite simply because humans are not immortal. Instead, we assumed \texttt{TIME\_start} to be within 60 years of the start of the time period (details in Web Appendix A.3). 

Now, we prepared to fit the disease progression model from Section~\ref{data:model}. Because symptoms were expected to worsen near diagnosis, time to diagnosis (in years) was a key covariate. Since cUHDRS at the last visit was our outcome, we defined time to diagnosis from the end of the time period, too. For uncensored subjects, \texttt{TIME\_end} was computed by subtracting their last visit date from their \texttt{DATE} of diagnosis. For censored subjects, \texttt{TIME\_end} was computed by subtracting their last visit date from the imputed $\widehat{\texttt{DATE}}$ of diagnosis instead, where $\widehat{\texttt{DATE}}$ was found by adding their conditional mean ${\rm E}(\texttt{TIME\_start}|\texttt{TIME\_start} > \texttt{FOLLOW\_UP}, \texttt{AGE}, \texttt{CAG})$ to their first visit date. 

\subsection{Strategic Recruitment for a Clinical Trial}\label{data:recruit}

Like the densities of time to diagnosis (Supplemental Figures~S9 and S10), the two imputation approaches led to different disease progression models, each with its own clinical implications (Table~\ref{table:predict-results}). We focused on adopting the models to guide recruitment for a new clinical trial in the following way. 

Suppose we were recruiting 200 at-risk subjects from their last regular study visit and that the clinical trial was expected to last for 2 years. Our recruitment strategy proceeds in two steps: (i) computing the subject-specific expected change in cUHDRS over the course of the clinical trial period (i.e., between recruitment and trial end 2 years later) and (ii) prioritizing subjects with the steepest expected drops in cUHDRS during that time. For demonstration, we begin by estimating one subject's symptom progression during the trial and discussing their resulting priority (Section~\ref{howto_one}) and then outline our large-scale recruitment strategy for an entire clinical trial (Section~\ref{howto_whole}). 

\begin{table}
\caption{Huntington's disease symptom progression models in PREDICT-HD fit using normal linear regression after imputing censored \texttt{TIME\_end} from last visit to diagnosis with conditional means.
\label{table:predict-results}}
\centering
\resizebox{\columnwidth}{!}{
\begin{threeparttable}
\begin{tabular}{llrclrc}
\hline
\multicolumn{2}{c}{ } & \multicolumn{2}{c}{\textbf{Extrapolated Conditional}} && \multicolumn{2}{c}{\textbf{Non-Extrapolated Conditional }} \\
\multicolumn{2}{c}{ } & \multicolumn{2}{c}{\textbf{Mean Imputation}} && \multicolumn{2}{c}{\textbf{Mean Imputation}} \\
\cmidrule(l{3pt}r{3pt}){3-4} \cmidrule(l{3pt}r{3pt}){6-7}
\textbf{Coefficient} && \textbf{Estimate} & \textbf{95\% CI} && \textbf{Estimate} & \textbf{95\% CI} \\
\hline
Intercept && $21.680$ & ($20.571$, $22.790$) && $23.298$ & ($22.349$, $24.246$) \\
\texttt{TIME\_end} && $0.084$ & ($-0.013$, $0.181$) && $0.117$ & ($-0.032$, $0.266$) \\
\texttt{cUHDRS\_start} && $1.048$ & ($0.941$, $1.155$) && $0.961$ & ($0.861$, $1.061$) \\
\texttt{TIME\_end}$\times$\texttt{cUHDRS\_start} && $-0.024$ & ($-0.036$, $-0.011$) && $-0.019$ & ($-0.036$, $-0.002$) \\
\texttt{AGE} && $-0.021$ & ($-0.046$, $0.003$) && $0.012$ & ($-0.009$, $0.032$) \\
\texttt{CAG} && $-0.089$ & ($-0.166$, $-0.012$) && $-0.092$ & ($-0.160$, $-0.025$) \\
\texttt{AGE}$\times$\texttt{CAG} && $0.006$ & ($0.001$, $0.011$) && $-0.014$ & ($-0.018$, $-0.010$)  \\
\bottomrule
\end{tabular}
\begin{tablenotes}[flushleft]
\item{\em Note:} \textbf{95\% CI}: 95\% Wald-type confidence interval based on the sandwich standard errors
\end{tablenotes}
\end{threeparttable}
}
\end{table}

\subsubsection{How to Estimate Symptom Progression and Prioritize a Subject for Recruitment}\label{howto_one}

Consider a randomly selected subject whose cUHDRS was already seen to decline from $\texttt{cUHDRS\_start} = 15.9$ to $\texttt{cUHDRS\_end} = 13.3$ between their first to last visits in PREDICT-HD, a pre-trial change of $\Delta(\texttt{cUHDRS}) = - 2.6$. In planning a clinical trial, the subject's symptom change \textit{during} the trial was more of interest than their change before, but this quantity is unobservable at recruitment. Fortunately, estimating this change in cUHDRS during the trial can be a powerful alternative. Specifically, we can predict subjects' cUHDRS 2 years from recruitment 
using the fitted symptom progression models, plugging in their cUHDRS at recruitment for \texttt{cUHDRS\_start} to obtain their expected cUHDRS at trial end, $\widehat{\texttt{cUHDRS\_end}}$. Then, expected symptom change during the clinical trial can be calculated from this prediction as $\widehat{\Delta}(\texttt{cUHDRS}) = \widehat{\texttt{cUHDRS\_end}} - \texttt{cUHDRS\_start}$. Thus, $\widehat{\Delta}(\texttt{cUHDRS}) < 0$ would indicate that the subject's symptoms are expected to worsen, and values farther from $0$ are expected to worsen more severely.

\begin{figure}
    \centering
    \includegraphics[width=0.95\textwidth]{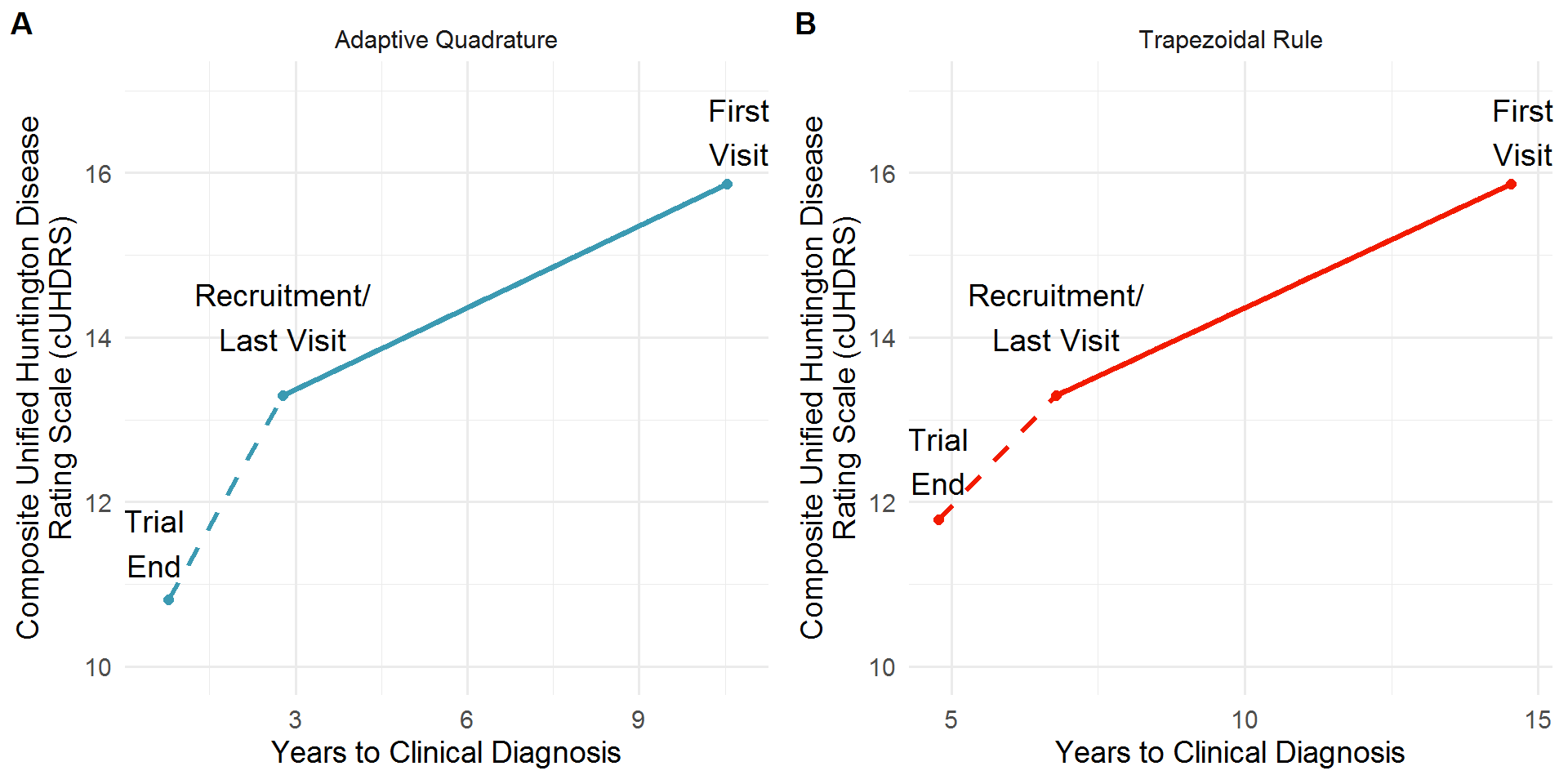}
    \caption{For each subject, we can estimate their cUHDRS at the end of the trial using the symptom progression models and then construct a complete trajectory of their symptom severity over study follow-up (i.e., the solid line from First Visit to Recruitment/Last Visit) and the 2-year clinical trial (i.e., the dashed line from Recruitment/Last Visit to Trial End).}
    \label{fig:trajectory}
\end{figure}

As a bonus, $\widehat{\texttt{cUHDRS\_end}}$ can also be used to construct a trajectory of the subject's symptom severity before and during the trial (Figure~\ref{fig:trajectory}). The quantities $\Delta(\texttt{cUHDRS})$ and $\widehat{\Delta}(\texttt{cUHDRS})$ summarize a subject's changes in symptoms before trial recruitment (observed change) and after trial recruitment (expected change), along this trajectory. For the same example subject, the model imputed using extrapolated conditional mean imputation predicted their cUHDRS to be $\widehat{\texttt{cUHDRS\_end}} = 10.8$ at the end of the trial, leading to an estimated change of $\widehat{\Delta}(\texttt{cUHDRS}) = -2.5$ during the trial. Based on this value, the subject had the 43rd largest estimated decrease in cUHDRS among censored subjects, making them high priority for recruitment. In contrast, the model imputed using non-extrapolated conditional mean imputation predicted their cUHDRS to change from $\texttt{cUHDRS\_start} = 13.3$ at recruitment to $\widehat{\texttt{cUHDRS\_end}} = 11.8$ at trial end for a smaller change of $\widehat{\Delta}(\texttt{cUHDRS}) = -1.5$, ranking 201st and giving this subject low priority for recruitment into a trial of $200$ subjects.

Because we saw in the simulation studies (Section~\ref{sims:compare_approaches}) that the non-extrapolated conditional mean imputation model estimates can be biased, particularly under extra heavy censoring rates like the 75\% in PREDICT-HD, we have more trust in the model imputed using extrapolated conditional mean imputation and believe that its expected symptom change of $\widehat{\Delta}(\texttt{cUHDRS}) = -2.5$ would be closer to the true one. In general, incorrectly prioritizing trial candidates (e.g., by mistakenly ranking someone 201st due to a biased model when they should really have been 43rd) means that non-ideal subjects may take spots away from others with potentially more to gain.

\subsubsection{How to Prioritize the Entire Study for Recruitment}\label{howto_whole}

We used the same process outlined above for everyone and then ordered the entire study by their estimated change in symptoms, $\widehat{\Delta}(\texttt{cUHDRS})$, starting from the biggest decline in function (i.e., largest decrease in cUHDRS). Then, we recruited subjects ranked 1--200, prioritizing subjects expected to have the worst symptom progression and with potentially the most to gain. We call this rank-based recruitment.

\begin{figure}[ht]
    \centering
    \includegraphics[width=0.95\textwidth]{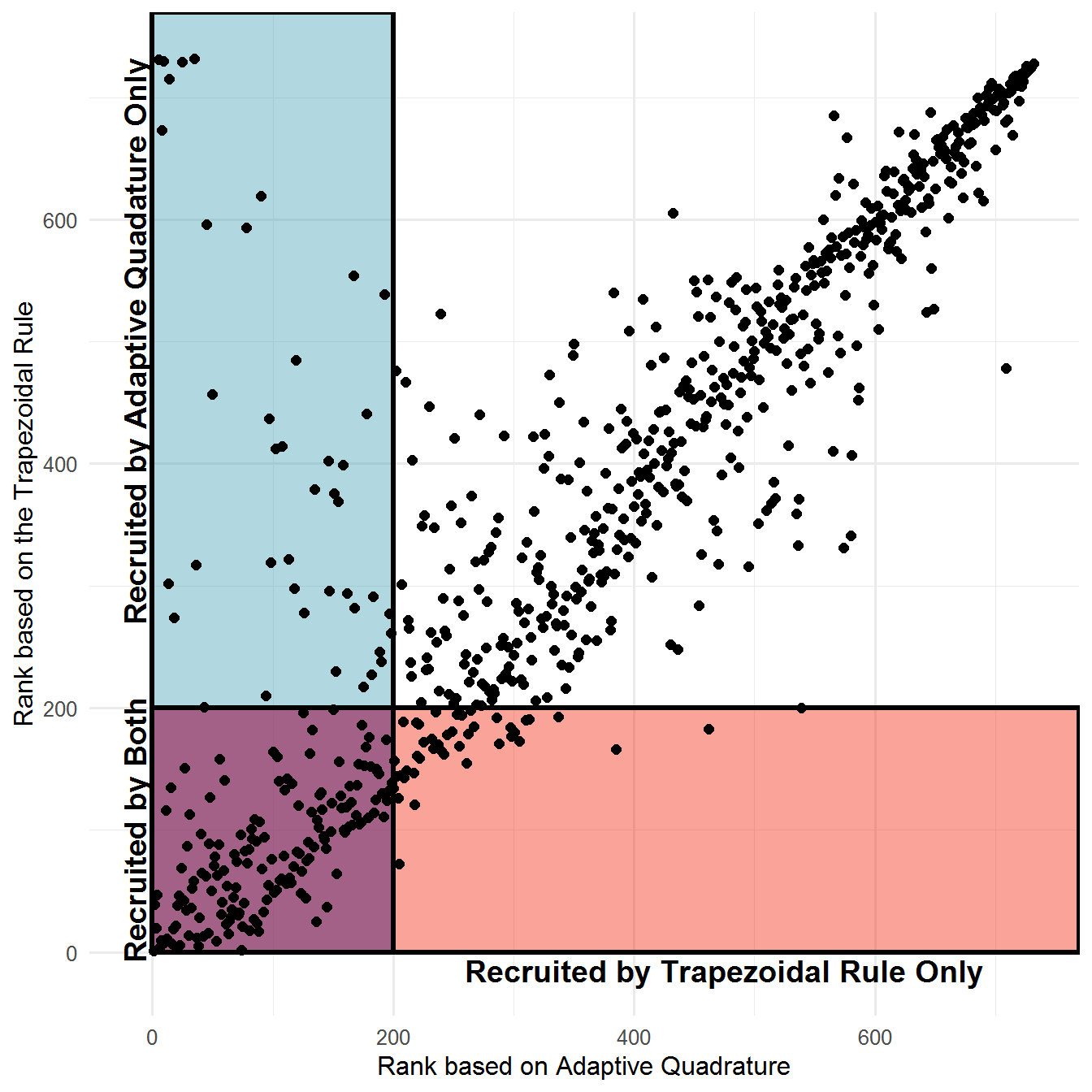}
    \caption{Subjects were ranked by their estimated changes in symptoms based on the disease progression models, starting from the biggest decline in function (i.e., largest decrease in cUHDRS), and the first $200$ subjects were subsequently recruited into the hypothetical clinical trial. The shaded regions capture subjects who would have been recruited based on each model, with the overlapping area in the lower left capturing subjects who would have been recruited based on either model. Points represent the $n = 732$ censored subjects from PREDICT-HD.}
    \label{fig:recruit}
\end{figure}

Although the PREDICT-HD study is over, we demonstrated our recruitment strategy with its data. Figure~\ref{fig:recruit} summarizes the recruitment statuses based on both disease progression models for the 732 censored subjects from the study. To introduce some realistic variability, we also created $1000$ new datasets of $732$ subjects each by resampling with replacement from the $732$ censored subjects in PREDICT-HD. In each resampled dataset, we applied our rank-based recruitment strategy twice: once with each disease progression model (extrapolated and non-extrapolated conditional mean imputation). On average, the models agreed on $158$ and $490$ subjects to recruit and not recruit, respectively. For the other $42$ subjects, the models disagreed, with non-extrapolated conditional mean imputation ``throwing away'' $42$ trial spots on subjects that the extrapolated model expected to have lesser changes in symptoms. For a summary across all resampled datasets, see Supplemental Figure~S11.

In an all-knowing world, we would recruit subjects for a new clinical trial who would have the steepest change in their symptoms without treatment to clearly measure the treatment effect (i.e., for a more obvious reduction in symptoms). However, we cannot know which subjects will have the steepest change in symptoms, so this is not a reasonable strategy. Recruiting subjects expected to have the steepest changes in symptoms is, though. With extrapolated and non-extrapolated conditional mean imputation, we modeled the progression of Huntington's disease symptoms, despite censoring in time to diagnosis, and used these models to guide recruitment for a hypothetical trial. The models disagreed on more than 20\% of who to recruit, but given its demonstrated accuracy in the simulations, we believe that using extrapolated conditional mean imputation will give statisticians confidence in their model and clinicians confidence in who they recruit based on it.

\section{Discussion}\label{sec:discuss}

We demonstrate through simulations that approximating the integral under Breslow's estimator with the trapezoidal rule, i.e., non-extrapolated conditional mean imputation, makes existing approaches miscalculate conditional means and leads to biased statistical inference. We propose a hybrid approach of extrapolating before imputing that can substantially reduce this bias. The proposed approach (i) combines the semiparametric Breslow's estimator with a parametric Weibull extension, before (ii) using adaptive quadrature to more completely approximate the integral up to infinity. 
Before recommending the Weibull extension, we provide an in-depth empirical investigation of how best to extend Breslow's estimator for integration to infinity, offering recommendations in various real-world settings. We then demonstrate how well our method corrects for the bias with the existing non-extrapolated conditional mean imputation approach, 
offering reduced bias in statistical inference from censored covariates through imputation. Finally, we applied our proposed methods to model the progression of Huntington's disease symptoms in the PREDICT-HD study relative to time of diagnosis, a censored covariate, and discussed using this model to guide recruitment for a new clinical trial. 

In our simulations and real-data analysis, we focused on linear regression modeling. However, the methods apply for any outcome model that captures the associations between $Y$, censored $X$, and $\bZ$. This flexibility is one of the strengths of imputation: Once the censored covariates are imputed with their conditional means, we can apply any of the usual modeling approaches. However, consistency for the conditional mean imputation estimators cannot be guaranteed in non-linear outcome models, like logistic regression \citep{Bernhardtetal2015}. 
 
Our proposed recruitment strategy takes a granular approach to targeting high priority subjects. Other strategies randomly sample from strata defined by a proxy for time to diagnosis. For example, \citet{Paulsen2019} create ``low'' and ``high'' risk groups from the CAP score \citep{Zhang2011}, where the high risk group is made up of subjects with CAP $>390.4$ who are believed to be nearest to diagnosis. One potential drawback of stratified strategies like this is that creating categories loses information from the continuous CAP variable. In other words, once subjects are placed into categories, there is no way for clinicians to gauge the relative priority of subjects within a risk group. For example, a subject with a CAP of $666.4$ (the largest in the study) has the same chance of being recruited as one with a CAP of $390.5$ (barely qualifying as high risk). In ranking subjects from smallest to largest expected symptom change rather than categorizing, our strategy empowers clinicians to directly recruit the highest priority subjects.

Even with our improvements, there are limitations to conditional mean imputation. 
Some bias remained when imputing with the extrapolated conditional mean imputation. 
Further investigation is needed to determine which survival function estimator to use for imputation, particularly under heavy and extra heavy censoring. In these high-censoring settings, a more structured parametric estimator might be preferred. Also, semiparametric imputation approaches like this one are sensitive to non-proportional hazards because they rely on the Cox model to estimate the survival function. We could test for this and modify the imputation model (e.g., with time-varying coefficients) to accommodate non-proportionality. Still, an entirely unspecified estimator, like the Kaplan--Meier, would be ideal if the data can support it. Finally, standard error estimation is problematic with single imputation approaches, like the one we discuss here, since the true variability for the model estimates is underestimated. However, the improvements we have proposed are needed for and could readily be adopted in a multiple imputation framework instead. 

There are several interesting statistical directions for future work. The first would be to extend our framework to capture multiple censored covariates. \citet{AtemMatsoukaZimmern2019} propose such an approach but use the trapezoidal rule to calculate the conditional means from the non-extrapolated survival function. 
Also, to our knowledge, imputation for randomly left-censored covariates has been thus far unaddressed and should be a relatively straightforward adaptation; the formula for the appropriate conditional means, ${\rm E}(X|X<W_i, \bZ_i)$, would need to be derived. 
There are also natural connections to methods other than imputation that require improper integration over a nonparametric or semiparametric survival estimator, for example, estimating mean residual life or maximum likelihood estimation with a censored covariate. Finally, an interesting clinical direction for future work might involve adopting our rank-based recruitment strategy for other measures of symptom progression (e.g., by ranking subjects on a proxy like CAP score). 



\section*{Acknowledgements}
The authors thank PREDICT-HD for permission to present their data. 

\section*{Supplementary Materials}
\begin{itemize}
    \item \textbf{Additional appendices, tables, and figures:} The Web Appendices and Supplemental Figures and Tables referenced in Sections 2--4 can be found in the Supplementary Materials online at \url{https://github.com/sarahlotspeich/hybridCMI/blob/main/supp.pdf}.
    \item \textbf{R-package for conditional mean imputation:} R-package \pkg{imputeCensRd} containing code to perform the imputation methods described in the article can be found at \url{https://github.com/sarahlotspeich/imputeCensRd}.
    \item \textbf{R code for simulation studies:}  R scripts to reproduce all simulations, tables, and figures, along with all simulated data, are available on GitHub at \url{https://github.com/sarahlotspeich/hybridCMI}.
\end{itemize}

\bibliographystyle{unsrtnat}
\bibliography{references}  

\end{document}